\documentclass[journal]{IEEEtran}

\ifCLASSINFOpdf
\else
   \usepackage[dvips]{graphicx}
\fi

\usepackage{url}

\usepackage{graphicx,amsmath}
\usepackage[T1]{fontenc}
\usepackage{tipa}
\usepackage{CJKutf8}

\usepackage{textcomp}
\usepackage{amssymb}
\usepackage{cite}
\usepackage{tabularx,booktabs,multirow}
\usepackage{eqparbox}
\usepackage{textcomp}
\usepackage[dvipsnames]{xcolor}
\usepackage{dblfloatfix}

\newcommand\ang[1]{\textlangle #1\textrangle}
\newcommand\mono{\texttt}
\newcommand\rar{\textrightarrow}
\newcommand\ipa[1]{/\textipa{#1}/}
\newcommand\dblpipe{\textipa{\textdoublepipe}}

\newcolumntype{C}{>{\centering\arraybackslash}X}
\newcommand{\hangline}{
    \par
    \hspace*{5pt}
    \ignorespaces
}
\newcommand{\breaklastline}[2]{{\setlength{\parfillskip}{0pt} #1 \par}{\noindent #2}}

\begin{document}

\title{PRESENT: Zero-Shot Text-to-Prosody Control}

\author{%
    Perry Lam, Huayun Zhang, Nancy F. Chen, Berrak Sisman, Dorien Herremans
    \thanks{Submitted on 16 Aug, 2024.}
    \thanks{Perry Lam and Dorien Herremans are with Singapore University of Technology \& Design, Singapore 487372 (e-mail: \protect\url{perry_lam@mymail.sutd.edu.sg} and \protect\url{dorien_herremans@sutd.edu.sg}).}
    \thanks{Huayun Zhang and Nancy F. Chen are with the Institute of Infocomm Research, A*STAR, Singapore 138632 (e-mail: \protect\url{Zhang_Huayun@i2r.a-star.edu.sg} and \protect\url{nfychen@i2r.a-star.edu.sg}).}
    \thanks{Berrak Sisman is with the University of Texas at Dallas, Richardson, TX 75080 USA (e-mail: \protect\url{berrak.sisman@utdallas.edu}).}
}

\maketitle

\begin{abstract}
Current strategies for achieving fine-grained prosody control in speech synthesis entail extracting additional style embeddings or adopting more complex architectures. To enable zero-shot application of pretrained text-to-speech (TTS) models, we present PRESENT (PRosody Editing without Style Embeddings or New Training), which exploits explicit prosody prediction in FastSpeech2-based models by modifying the inference process directly. We apply our text-to-prosody framework to zero-shot language transfer using a JETS model exclusively trained on English LJSpeech data. We obtain character error rates (CER) of 12.8\%, 18.7\% and 5.9\% for German, Hungarian and Spanish respectively, beating the previous state-of-the-art CER by over 2$\times$ for all three languages. Furthermore, we allow subphoneme-level control, a first in this field. To evaluate its effectiveness, we show that PRESENT can improve the prosody of questions, and use it to generate Mandarin, a tonal language where vowel pitch varies at subphoneme level. We attain 25.3\% hanzi CER and 13.0\% pinyin CER with the JETS model. 
All our code and audio samples\footnote{\url{https://github.com/iamanigeeit/present} and \url{https://present2023.web.app/}} are available online. 
\end{abstract}


\IEEEpeerreviewmaketitle

\vspace{-7pt}

\section{INTRODUCTION}
\label{sec:intro}

\IEEEPARstart{R}{ecent} neural text-to-speech (TTS) models have approached human-like naturalness in read speech. However, attaining similar expressiveness levels remains a challenge. A growing body of research aims to add and control speech prosody variations, progressing from digital signal processing (DSP) methods to style and emotion embeddings built into TTS architectures or even entire models to extract and transfer prosody.

On the waveform level, prosody control can be achieved through operations like time-stretching and pitch-shifting. DSP methods such as TD-PSOLA \cite{TD-PSOLA} and WORLD \cite{WORLD}, despite their known artifacts, are still widely applied due to their speed and ease of use. Remarkably, they can perform as effectively as neural approaches like Controllable LPCNet \cite{CLPCNet}.

In contrast, expressive TTS systems \cite{expressive-tts} allow the user to specify a style or emotion label during inference. 
Recent TTS models incorporate style or emotion information by extracting a reference embedding that represents the prosody or emotion from labelled audio, and adding it to the model encoder. This can be combined with a style bank for smooth style variation, such as in Global Style Tokens \cite{gst}. Further extensions include phoneme-level prosody control and hierarchical autoencoders to ensure coherence over the whole utterance \cite{chive}. 



All of these approaches, however, require extra model components and/or further training. Therefore, to combine the simplicity of DSP methods with the naturalness of neural speech generation, we empower users to directly control prosody using the input text and inference parameters without the need for any fine-tuning or architectural modifications. We contribute significantly in the following three areas:
\begin{itemize}
    \item \textbf{Extraction of prosodic effects from text}, such as extended duration in ``A looooong time" or the intonation variations in questions like ``What was that?". We take these prosodic parameters and modify the inference method of any TTS model with explicit duration, pitch, and energy (DPE) predictions to generate varying speech.
    \item \textbf{Zero-shot language transfer} with no target-language audio, relying solely on linguistic knowledge and modifying DPE to create new phonemes and speech patterns.
    \item  \textbf{Subphoneme-level control}, achieved by subdividing phonemes and applying custom pitch and energy over the subdivisions, which helps us change long vowel intonation and model tonal languages like Mandarin.
    
\end{itemize}
Though our primary goal is to explore the limits of editing inference-time prosody predictions, in doing so, we achieve state-of-the-art results in zero-shot language transfer.


The rest of this paper is organized as follows: Section 2 summarizes relevant research, Section 3 describes our approach, Section 4 lists our experiment results and Section 5 concludes our paper.

\begin{figure*}[!b]
    \vspace{-10pt}
    \centering
    \includegraphics[width=\textwidth]{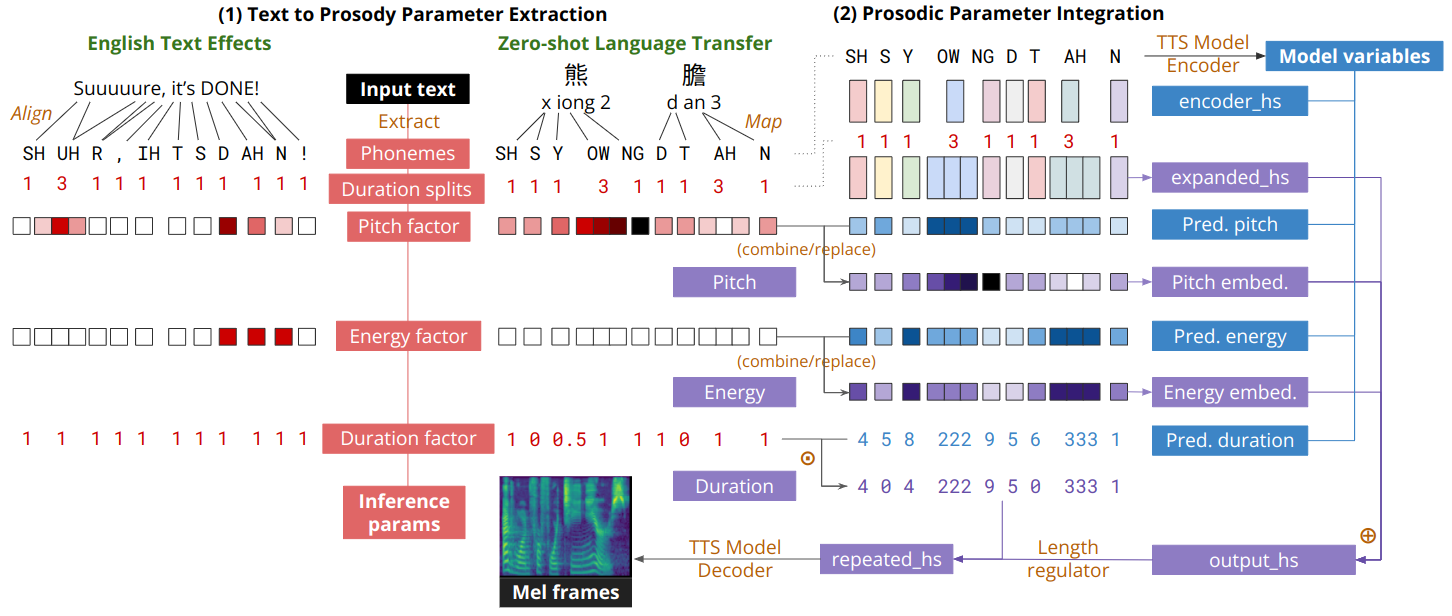}
    \setlength{\abovecaptionskip}{-14pt}
    \caption{Overview of inference modification in PRESENT. \textcolor{Green}{Green} items are tasks, \textcolor{Black}{black} items are input/outputs, \textcolor{Red}{Red} items represent PRESENT outputs, \textcolor{RoyalBlue}{blue} items are the outputs from the FS2-based model, \textcolor{Purple}{purple} are the combined outputs, and \textcolor{brown}{gold} are operations.}
    \label{fig:diagram}
\end{figure*}

\vspace{-5pt}

\section{RELATED WORK}
\label{sec:related}

Based on our main contributions, we divide the related work into the broad categories of (1) speech effect tagging, (2) zero-shot language transfer, and (3) fine-grained prosody control.

\vspace{-5pt}

\subsection{Speech Effects Tagging}
Text-based methods for manipulating speech can be categorized into explicit and implicit forms. Explicit speech descriptors such as gender and emphasis have been integrated into the industry standard Speech Synthesis Markup Language (SSML) over the past two decades \cite{SSML}. 
Yet, there has been relatively limited published research on SSML, even though there have been notable introductions of TTS models with new style tags, as demonstrated in \cite{style-tag}.


Implicit methods 
establish connections between prosodic features and text, such that a sentence like "this is interesting!" would sound excited. Typically, this means that the text embeddings from a language model 
are used as input either at the subword \cite{cauliflow} \cite{speech-bert} or phoneme level \cite{phoneme-bert} \cite{png-bert}.  However, due to their inherent limitations in customizing prosody changes, recent projects inspired by advancements in computer vision and language processing let user input a natural-language style prompt like ``sighing with helpless feeling" to generate prosodic output \cite{instruct-tts}.



\vspace{-10pt}

\subsection{Zero-Shot Language Transfer}

While multilingual TTS models have existed for some time, they rely on large multilingual corpora, which disadvantages lower-resourced languages. Transfer learning \cite{meta-learning} \cite{low-resource-transfer} and data balancing \cite{universal-tts} techniques have been employed, but these still require at least some audio data. With only International Phonetic Alphabet (IPA) transcriptions in the target language, \cite{phone-features-0-shot} proposed using IPA phonological features to extend existing models on unseen phonemes, whereas two very recent large models have proposed zero-shot TTS with only text data available in the target language.

The first model, VALL-E X \cite{VALL-E-X}, uses AudioLM \cite{audio-lm} codec codes as acoustic tokens in place of mel spectrograms as intermediate features, and treats the cross-lingual TTS model as a massive language model (LLM) that can be trained with self-supervised masking. Given a speech sample in the source language, plus source and target language phoneme sequences, it extracts the source acoustic tokens from the speech sample and the LM predicts the target acoustic tokens. Since the acoustic tokens contain speaker, recording conditions, and subphoneme information, the decoder can reconstruct the waveform for the target language in the source speaker's voice.

The second model, ZM-Text-TTS \cite{zero-shot-tts}, also uses masked multilingual training, but on IPA / byte tokens and raw text. The pretraining results in a language-aware embedding layer that is fed to a conventional multilingual TTS system for training with seen languages, and the model can accept IPA / byte tokens for unseen languages during inference. Nevertheless, VALL-E X is not publicly available, and ZM-Text-TTS does not account for prosody in language transfer.

\vspace{-5pt}

\subsection{Fine-grained Prosody Control}

\breaklastline{
As utterance-level styles are now commonplace, research has shifted to controlling prosody at the phoneme level. Since acceptable prosodies are obtained by learning and sampling from a variational latent space, hierarchical variational auto-encoders (VAEs) \cite{hierarchical-vae} can achieve fine prosodic gradations,}{down to the syllable, phone or even frame level \cite{chive}.}

Alternatively, others use phone-level DPE for interpretable prosody control. This was the approach of earlier research~\cite{robust-prosody}, but to improve output naturalness, \cite{prosody-clustering} and \cite{prosody-tokens} used k-means clustering on duration and pitch for each speaker, and kept the resulting centroids as discrete prosody tokens. This allows the tokens to be substituted at inference time to customize prosody, while decoding with a prosody attention module ensures information flows to the output. Meanwhile, since the advent of explicit DPE models like FastSpeech2 \cite{fastspeech2}, models like \cite{emotional-prosody} and \cite{empathic} have extra modules attached that accept emotional dimensions (valence, arousal, dominance) that feed into phone-level DPE predictors, allowing for continuous emotion control.

\vspace{-3pt}

\section{PROPOSED METHOD}
\label{sec:method}
PRESENT offers a versatile approach to (1) extract inference parameters and (2) integrate them with explicit DPE predictions to generate variations in pronunciation and prosody, all without requiring additional modules or fine-tuning. The specific method of parameter extraction and integration is adaptable to the task at hand and can be customized by the user. Fig \ref{fig:diagram} illustrates this process with examples for English text-to-prosody and English-to-Mandarin language transfer.



\vspace{-10pt}
\subsection{Obtaining Prosodic Effects from Text}

We preprocess the input text to capture common dialogue features such as CAPS or *asterisks* for emphasis, repeaaaaated letters or ti\textasciitilde\textasciitilde ldes for long phonemes, and special characters like underscores and carets or questions for tone modification. As TTS systems usually rely on phoneme input, we align the text to phonemes so that DPE changes can be applied at the right positions. While grapheme-to-phoneme (G2P) systems are widely available, G2P alignment systems are outdated and lack Python implementations. Thus, we develop our own aligner combining the ``Phonetic Alignment" and ``IP Alignment" approaches in \cite{alignment}. We begin with a list of allowed grapheme-phoneme mappings (e.g. \mono{ch} \rar \mono{\{CH,K,SH\}}) and search for a constraint-satisfying alignment, or return the least-cost path by dynamic programming if an alignment cannot be found. For example, given the word \mono{where} and pronunciation \mono{W EH R}, the aligner returns \mono{wh}\rar \mono{W}, \mono{e}\rar \mono{EH}, \mono{r}\rar \mono{R}, \mono{e}\rar $\varnothing$. With an invalid pair like \mono{whence} and \mono{W Z EH T}, the aligner chooses the minimal cost of disallowed mappings: $\varnothing$\rar \mono{Z}, \mono{n}\rar \mono{T}, \mono{c}\rar $\varnothing$. The aligner also allows us to detect possibly wrong dictionary entries, such as the \mono{\textbf{G}} of \mono{EEG}\rar\mono{IY IY \textbf{G} IY} in the CMU Pronouncing Dictionary, and correct them (\mono{IY IY \textbf{JH} IY}). 

\vspace{-4pt}

\subsection{Generating Parameters for Zero-Shot Language Transfer}

Manipulating DPE directly allows us to create phonemes and intonation patterns not found in our model's language. For instance, we can approximate the German and Spanish /x/ with ARPAbet [HH K HH] and durations [1, 0, 1] to velarize [HH] without producing a distinct [K]. Since ZM-Text-TTS tested on German (de), Hungarian (hu), and Spanish (es), we describe the main ideas for adapting American English models to these languages in Table \ref{tab:css10_map}, and also include Mandarin (cmn) for the next section. The full conversion tables are in our code.

\begin{table}[h]
\vspace{-7pt}
\caption{Mapping rules to ARPAbet. IPA symbols in /slashes/, pinyin \ang{angle brackets}, ARPAbet [square brackets]. D = duration factor, P = pitch change, E = energy change, \dblpipe\ = word sep.}
\setlength{\tabcolsep}{3pt}
\begin{tabularx}{\columnwidth}{c|>{\raggedright\arraybackslash}X}
\toprule
All &     Use combinations, possibly with zero duration, for phonemes that don't exist in English (for \ipa{B L \textltailn\ ç x y \o} etc). Examples: \hangline{
        German \ipa{\oe}\rar [W EH] D=[0,1] and \ipa{ç}\rar [H SH S] D=[0,1,0]
    } \hangline {
        Hungarian \ipa{y}\rar [UH Y] D=[0,1] and \ipa{\textbardotlessj}\rar [G Y] D=[0.7,0]
    } \hangline {
        Spanish \ipa{r}\rar [R HH R] D=[1,0,1] and \ipa{B}\rar [B V] D=[0,1]
    } \hangline {
        Mandarin \ang{zh}\rar [T SH] D=[1,0] and \ang{x}\rar [SH S] D=[1,0]
    }
\\ \midrule
de &
    Shorten long vowels corresponding to German short vowels: \hangline{
        \ipa{A}\rar [AA] D=0.5
    }
    
    Make schwa clearer and prevent merging into next phoneme: \hangline{
        \ipa{@}\rar [AX] D=1.5 E=+1 and [AH \dblpipe]\rar [AH ,] D=[1,0]
    }
\\ \midrule
hu &
    Shorten long vowels corresponding to Hungarian short vowels: \hangline{
        \ipa{u}\rar [UW] D=0.5
    }
    
    Reduce phoneme lengths as Hungarian has faster speaking speed: \hangline{
        \ipa{b}\rar [B] D=0.7
    }
    
    Double consonants for long consonants: \hangline{
        \ipa{k:}\rar [K K] D=[0.7,0.7]
    }
\\ \midrule
es &
    Reduce phoneme lengths as Spanish has faster speaking speed: \hangline{
        \ipa{t}\rar [T] D=0.7
    }
    
    Shorten long vowels corresponding to Spanish short vowels: \hangline{
        \ipa{o}\rar [OW] D=0.4
    }
    
    Insert semivowels or very short pauses between consecutive vowels to avoid diphthongization, and raise the stressed vowel: \hangline{
        \ipa{o."i}\rar [OW W IY] D=[0.4,0.4,0.7] P=[0,0,+1], E=[0,0,+0.5]
    }
    
    Double plosives between vowels to prevent devoicing: \hangline{
        \ipa{apa}\rar [AA P P AA] D=[0.7, 0, 0.7, 0.7]
    }
\\ \midrule
cmn & 
    Define conversion for initial and rimes instead of individual phonemes, as they don't always combine sequentially: \hangline {
        \ang{i}\rar [IY ,] D=[1,0] but \ang{in}\rar [IH IY N] D=[1,0,1]
    }
    
    Use 0-duration voiceless phone to stop voicing vs [HH] for aspiration: \hangline{
        \ang{g}\rar [G K] D=[1,0] and \ang{k}\rar [K HH] D=[1,0.5]
    }
    
    Syllables that are pronounced differently from initial + rime mapping have their own rules: \hangline{
        \ang{i} before \ang{z(h),c(h),s(h)} \rar [Z UH] D=[0.5,0.7] but
    } \hangline{
        \ang{chi}\rar [CH HH R R] D=[1,0.5,1,1]
    }
    
    Add pause before characters that start with a vowel: \hangline{
        \ang{ai}\rar \ipa{Pai}\rar [, AY] D=[0.2,1]
    }
    
    Set glides to half duration: \hangline{
        \ang{iu}\rar [Y OW] D=[0.5,1]
    }
\\ \bottomrule
\end{tabularx}
\label{tab:css10_map}

\vspace{-6pt}
\end{table}

\subsection{Subphoneme-level Control and Tonal Languages}
To achieve tone contour effects like the rising-falling pitch in ``Suuuuure!", the phoneme must be split and each subphoneme assigned separate pitches. Thus, we repeat the encoder output $\boldsymbol{h}$ for the divided phoneme (boxes for \mono{OW} and \mono{AH} of \mono{encoder\_hs} in Fig \ref{fig:diagram}). This differs from simply repeating the phonemes for inference, as that would generate varying $\boldsymbol{h}$ and possibly make the phoneme pronounced multiple times.

One evident use case for pitch effects pertains to questions. While humans can clearly perceive a question via prosody, TTS systems still lack proper intonation. For English question prosody, 
\cite{into-tts} enabled users to choose from a range of pretrained prosody templates. However, to maintain simplicity and avoid adding the complexity of language models, we follow the prosodic analysis of \cite{question-prosody}, applying a low-to-high accent on the locus of interrogation and the final word of the question to convey question intonation.

Another critical test of our subphoneme tone contour approach is its ability to model tonal languages. 
After applying the phoneme changes in Table \ref{tab:css10_map}, we split each vowel nucleus into subphonemes and assign their pitch following Mandarin tone contours in Table \ref{tab:tone-pitch}. Contours on the five-point tonal scale are then normalized to pitch values between $[-2.0, +2.0]$. 


\begin{table}[h]
 \vspace{-10pt}
\centering
\caption{Tone-pitch mapping.}
\begin{tabular}{c|c|c|c|c|c}
\textbf{Tone}    & 1      & 2      & 3          & 4      & 5 \\
\specialrule{1pt}{1pt}{1pt}
\textbf{Contour} & 55     & 24     & 212        & 52     & - \\
\textbf{Pitch}   & +2, +2 & -1, +1 & -1, -2, -1 & +2, -1 & 0 \\
\end{tabular}
\label{tab:tone-pitch}
\end{table}
\vspace{-5pt}


The initial and coda (if any) take the start and end pitch of the tone contour. Table \ref{tab:arpa-pinyin} demonstrates one example of how pinyin \ang{tian2}\ maps to [T HH Y EH N]. Pitch transitions are smoothed across syllables to avoid abrupt pitch changes.

\begingroup
\setlength{\tabcolsep}{4pt}
\begin{table}[h]
 \vspace{-9pt}
\centering
\caption{Example of ARPA-pinyin mapping.}
\begin{tabular}{r|c|c|c|ccc|c}
\textbf{ARPA}        & T  & HH & Y    & \multicolumn{3}{c|}{EH}                                      & N  \\
\textbf{Duration}    & \texttimes1  & \texttimes0.5 & \texttimes0.5    & \multicolumn{3}{c|}{\texttimes1}                                      & \texttimes1  \\
\textbf{Subphonemes} & T  & HH & Y    & \multicolumn{1}{c|}{EH}   & \multicolumn{1}{c|}{EH}   & EH   & N  \\
\textbf{Pitch}       & -1 & -1 & -1 & \multicolumn{1}{c|}{-0.33} & \multicolumn{1}{c|}{0.33} & +1 & +1
\end{tabular}
 \vspace{-5pt}
\label{tab:arpa-pinyin}
\end{table}
\endgroup

As Mandarin is a syllable-timed language, we keep the \texttimes1 duration 
constant, with the neutral tone at half duration. Finally, we leverage \mono{pywordseg} to segment Mandarin text and introduce brief pauses between words for better enunciation. 
\vspace{-7pt}

\section{EXPERIMENTS}
\label{sec:experiments}

We conducted our experiments using the ESPnet \cite{espnet-tts} toolkit for reproducibility. Our source model was the publicly released English-only single-speaker JETS \cite{jets} model pretrained on LJSpeech, known for achieving state-of-the-art naturalness.

\vspace{-11pt}
\subsection{Zero-Shot Language Transfer}
\vspace{-1pt}
We first evaluate the ability of the PRESENT to produce intelligible speech in other languages. To generate audio samples, we extract raw text from the CSS10 dataset and phonemize them with \mono{espeak-ng}, then perform phoneme conversion and DPE editing. As a baseline, we employ ZM-Text-TTS \cite{zero-shot-tts}, the only open-source zero-shot language transfer system. As ZM-Text-TTS was trained and evaluated on the CSS10 \cite{css10} datasets' European languages subset, we compare PRESENT on the 3 languages on which ZM-Text-TTS has done zero-shot TTS: German, Hungarian, and Spanish. 

For fair comparison, we follow them in evaluating character error rate (CER) by running generated audio through Whisper's \cite{whisper} multilingual speech recognition. We use the large-v2 model from SYSTRAN's faster-whisper 
on default settings for its speed and robustness. CER is computed from the length-normalized Levenshtein distance between Whisper transcripts and ground truth, ignoring punctuation and whitespace.

Since ZM-Text-TTS pretrains on multilingual text before training on text-audio pairs, there are two settings in their evaluation: text-unseen (where the target language text is not available for pretraining) and text-seen (where the text is available, but there is no paired audio). Naturally, the text-unseen case leads to higher CER. 

\begin{table}[h]
\vspace{-8pt}
\caption{\label{tab:css10_cer} CER comparison. Euro. Lang. = European Languages. Ground Truth = raw CSS10 audio. Range given for Spanish ZM-Text-TTS is from using either phonemes or bytes as input.}
\begin{tabularx}{\columnwidth}{c|rccc}
\toprule
\textbf{Target}             & \textbf{(Source Languages) Model}
& \textbf{Text}             & \textbf{CER}
\\ \midrule
\multirow{4}{*}{German}     & \multirow{2}{*}{(6 Euro. langs.) ZM-Text-TTS}
& Unseen                    & 38.75 \\
                            & 
& Seen                      & 28.01 \\
                            & (English only) PRESENT
& Unseen                    & \textbf{12.82} \\
                            & Ground Truth
& --                        & 3.90 \\
\midrule
\multirow{4}{*}{Hungarian}  & \multirow{2}{*}{(6 Euro. langs.) ZM-Text-TTS}
& Unseen                    & 52.62 \\
                            &
& Seen                      & 50.11 \\
                            & (English only) PRESENT
& Unseen                    & \textbf{18.73} \\
                            & Ground Truth
& --                        & 3.15 \\
\midrule
\multirow{4}{*}{Spanish}    & \multirow{2}{*}{(7 Euro. langs.) ZM-Text-TTS}
& Unseen                    & 44.75 -- 64.07 \\
                            & 
& Seen                      & 11.69 -- 18.27 \\
                            & (English only) PRESENT
& Unseen                    & \textbf{5.92} \\
                            & Ground Truth
& --                        & 1.97 \\
\bottomrule
\end{tabularx}
\vspace{-6pt}
\end{table}

PRESENT reduces the previous state-of-the-art CER by over 2$\times$ for each language, even with a single off-the-shelf English-only model to generate them with no further training. In fact, our CER is close to the ZM-Text-TTS multilingual model trained \textit{with} target audio (German 9.76, Hungarian 9.11 and Spanish 5.32). This shows that phoneme conversion followed by prosody manipulation is critical for zero-shot language transfer.

\vspace{-10pt}
\subsection{Subphoneme-Level Control}


For question prosody, we took the first 10 dialogues from the DailyTalk dataset \cite{dailytalk} and extracted the first single-sentence question from each of them, making 10 questions in total. We report the Mean Opinion Score (MOS) for ground truth audio from DailyTalk, unaccented JETS-generated audio, and PRESENT-accented audio. Experiments were conducted with PsyToolkit \cite{psytoolkit-1} \cite{psytoolkit-2} and 15 responses were received.

\begin{table}[h]
\vspace{-8pt}
\centering
\caption{MOS for Question Prosody.}
\begin{tabular}{r|cc}
\toprule
\textbf{Ground Truth} & \textbf{JETS} & \textbf{PRESENT} \\
\midrule
4.46                  & 3.73          & \textbf{3.92}    \\
\bottomrule
\end{tabular}
\vspace{-5pt}
\label{tab:comparisons}
\end{table}


We then evaluate the ability of the JETS model to produce intelligible Mandarin speech by synthesizing speech based on the AISHELL-3 test set transcripts. As a baseline, we take the IPA multilingual model (pretrained on 7 European languages) model from ZM-Text-TTS,
convert pinyin transcripts to IPA, and use the best-approximation phoneme when a Mandarin phoneme does not exist in the pretrained IPA symbol set. 
We then input the the ground truth, PRESENT, and ZM-Text-TTS audio into the state-of-the-art Paraformer automatic speech recognition (ASR) framework \cite{paraformer}, and ensure transcriptions only contain Chinese by masking decoder outputs to $-\infty$ for alphanumeric tokens. Due to hallucination issues in individual models, we use a mixture-of-experts consisting of the aishell2-vocab5212 and paraformer-large-vocab8404 models.

The CER for transcriptions are computed at both Hanzi level and pinyin level in Table \ref{tab:mandarin_cer}. 
As Mandarin has many homophones, pinyin CER is a better measure of intelligibility; romanization also makes it comparable to European-language CER where a change like \mono{la} to \mono{le} is 50\% CER, not 100\%. Thus, we romanize Mandarin transcripts with \mono{pypinyin} and include pinyin CER with tone counting as one character (i.e. \mono{ping1} versus \mono{ping2} would be 20\% CER).

We also measured MOS (naturalness only) by selecting 15 utterances from the AISHELL-3 test set that made sense as full standalone sentences without jargon, rare words, or names. We skipped testing ZM-Text-TTS and PRESENT without tones or duration control, since they did not produce intelligible results. As before, the survey was created with PsyToolkit and 10 responses were received from Mandarin speakers.

\begin{table}[h]
\vspace{-8pt}
\centering
\caption{\label{tab:mandarin_cer} English-to-Mandarin language transfer results. 
}
\begin{tabular}{lcccc}
\toprule
\multicolumn{1}{c}{} & \% \textbf{Hanzi CER} & \% \textbf{Pinyin CER} & \textbf{MOS}
\\ \midrule
Ground Truth           & 1.2            & 0.9           & 4.65
\\ \midrule
PRESENT                & \textbf{25.3}  & \textbf{13.0} & 2.18 \\
-- w/o tones           & 59.5           & 33.8          & 1.92 \\
-- w/o tones/duration  & 105.4          & 63.9          & -       \\
ZM-Text-TTS            & 95.0           & 71.7          & -        \\
\bottomrule
\end{tabular}
 \vspace{-3pt}
\end{table}

The dramatic CER reductions from ZM-Text-TTS to PRESENT with phoneme conversion, duration and tones applied in succession demonstrates the effectiveness of our subphoneme-level DPE control. Using Latin orthography, the English-to-Mandarin language transfer CER is equal to the average of English-to-\{German, Hungarian, Spanish\}, and even outperforms all previous baselines on those languages. 
Still, MOS testing reveals the naturalness limitation of PRESENT-generated audio for human listeners due to the strong American accent, despite some improvement via tone contouring.





\vspace{-3pt}

\section{CONCLUSIONS}
We have introduced PRESENT, a novel approach that explores the limits of using only DPE predictions in a single-speaker English-only JETS model, without any additional embeddings or training. Our technique allows us to create prosodic effects from text and synthesize speech in unseen languages. 
Our zero-shot language transfer far outstrips previous state-of-the-art for European languages. Furthermore, the phoneme conversion and tone contour techniques we develop could enable direct accented speech generation (as the results are all American-accented), or TTS for hundreds of tonal minority languages within the Mainland Southeast Asian linguistic area
that are only recorded in phonetic transcriptions. 

\pagebreak
\bibliographystyle{ieeetran}
\bibliography{refs}

\end{document}